# Experimenting an optical second with strontium lattice clocks

R. Le Targat[1]*, L. Lorini[1,2], Y. Le Coq[1], M. Zawada[1,3], J. Guéna[1], M. Abgrall[1], M. Gurov[1], P. Rosenbusch[1], D. G. Rovera[1], B. Nagórny[1,3], R. Gartman[1,3], P. G. Westergaard[1], M. E. Tobar[4], M. Lours[1], G. Santarelli[1], A. Clairon[1], S. Bize[1], P. Laurent[1], P. Lemonde[1], J. Lodewyck[1]*

[1] LNE-SYRTE, CNRS, UPMC, Observatoire de Paris, 61 Avenue de l'Observatoire, 75014 Paris, France, [2] Istituto Nazionale di Ricerca Metrologica, Strada delle Cacce 91, 10135 Torino, Italy, [3] Institute of Physics, Nicolaus Copernicus University, Grudziadzka 5, PL-87-100 Toruń, Poland, [4] School of Physics, M013, The University of Western Australia, 35 Stirling Hwy, Crawley, 6009, WA, Australia.

* To whom correspondence should be addressed: rodolphe.letargat@obspm.fr, jerome.lodewyck@obspm.fr

**Progress in realizing the SI second had multiple technological impacts and enabled to further constraint theoretical models in fundamental physics. Caesium microwave fountains, realizing best the second according to its current definition with a relative uncertainty of $2\text{-}4\times10^{-16}$, have already been overtaken by atomic clocks referenced to an optical transition, both more stable and more accurate. Are we ready for a new definition of the second? Here we present an important step in this direction: our system of five clocks connects with an unprecedented consistency the optical and the microwave worlds. For the first time, two state-of-the-art strontium optical lattice clocks are proven to agree within their accuracy budget, with a total uncertainty of $1.5\times10^{-16}$. Their comparison with three independent caesium fountains shows a degree of reproducibility henceforth solely limited at the level of $3.1\times10^{-16}$ by the best realizations of the microwave-defined second.**

When the international system of units was formalized in 1960, the definition of the second was still derived from an astronomical reference, the Ephemeris Time (ET). Following the measurement[1] of the caesium ground state hyperfine F=3–F'=4 transition frequency (~9.19 GHz) in terms of the ET second, this transition was chosen as reference for a new definition of the SI second in 1967, in replacement of the ET second. Over the years, the realization of this unit progressed, and the best atomic fountains have now reached accuracies of $\sim 2\text{-}4\times10^{-16}$ in relative units[2]. In parallel, an operational architecture was developed to establish and distribute time scales based on this microwave second, e.g. via satellite systems like GPS or TWSTFT. The advent of the frequency combs[3,4] in 2000 led to a revolution in the field of frequency metrology. At once, this simple and accurate way to link the optical (~$10^{15}$ Hz) and the microwave domains made it possible to take full benefit from decades of progress in optical spectroscopy and laser cooling of atoms. It stimulated the development of a new generation of atomic clocks, based on optical transitions, and realizing their own unperturbed frequency with an uncertainty now surpassing the best realizations of the SI second. The use of the term accuracy is now often extended to quantify the control of systematic effects in optical clocks, even though they are not realizing the SI second.

A few optical clocks operated with single ions feature a record control of the systematics in the $10^{-17}$ range[5–7] or even $8.6\times10^{-18}$ (ref 8). Research on Optical Lattice Clocks (OLCs) is comparatively recent[9–13] but in less than 10 years, they have reached the $10^{-16}$ level[14–16] with excellent prospects towards the $10^{-17}$ level. The large number of probed atoms, on the order of $10^4$, already allows unprecedented statistical resolutions, or frequency stabilities, of a few $10^{-16}$ at one second, demonstrated in other groups[17–20]. The introduction of the magic wavelength[9,21] enables to minimize the impact of the trapping potential on the accuracy, at least down to $10^{-17}$ (ref 22). Despite the fermionic nature of atoms in most OLCs, residual atomic density effects in many−particle systems were highlighted[23–25] and proved to be controllable below $10^{-18}$ (ref 20). The parallel development of optical fibre links[26–28] allows the dissemination of an optical phase with negligible added noise over long distances. With such a link, the possibility of comparing two remote OLCs has been demonstrated[29] recently. These studies are preparing the ground for a possible redefinition of the second, and seven optical transitions have already been endorsed or recommended as secondary representations of the SI second. Nevertheless, a prerequisite before any of these candidates is chosen as new primary standard is to assess the reliability of their connexion to the current SI second.

In this article, we describe the experimentation of an optical second based on an ensemble of five clocks. Firstly, we compare two optical clocks of the same nature, which is the ultimate test to validate the asserted accuracy. For the first time, two OLCs with a state-of-the-art uncertainty of $1.1\times10^{-16}$ are proved to agree within their accuracy budget. Secondly, we demonstrate a reproducible link between these clocks and the microwave second realized by three atomic fountains. This comparison is henceforth limited by the accuracy of the caesium fountains and constitutes the precision limit with which any frequency can be measured with respect to the current SI second. This work thus reports on a comprehensive and consistent set of optical-optical and optical-microwave clock comparisons prototyping an optical second.

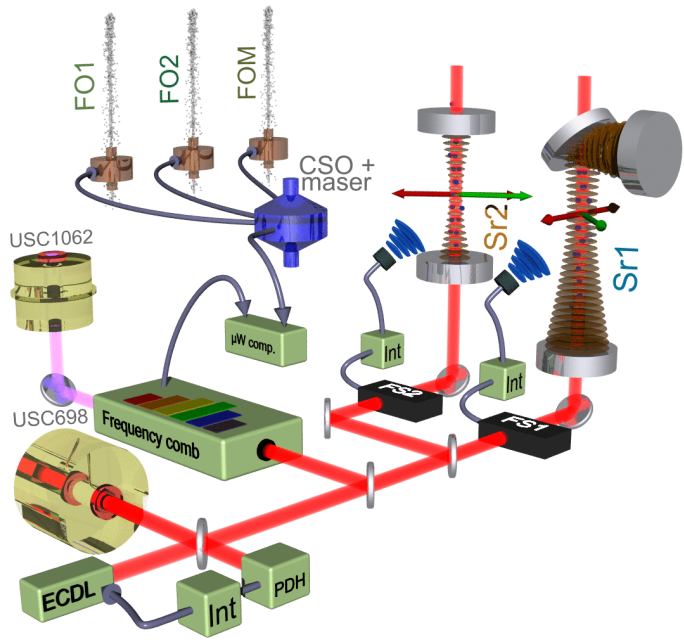

**Fig. 1: LNE-SYRTE optical to microwave measurement chain**. An Extended Cavity Laser Diode (ECDL) referenced to an Ultrastable Cavity (USC698) via a Pound-Drever-Hall (PDH) feedback loop (Int) probes the atoms trapped in the respective lattices of the two optical clocks (Sr1 and Sr2). Red arrows represent the lattice polarizations, green arrows the quantization axis. The detection systems keep the laser probes independently at resonance with the two clocks via a feedback (Int) on the Frequency Shifters (FS1 and FS2). This feedback is asynchronous, ensuring there is no common systematic effect due to the clock laser. The frequency of the USC698 is measured by a titanium-sapphire laser based frequency comb referenced to an auxiliary Ultrastable Cavity (USC1062). The division of the USC698 frequency to the microwave domain is compared (µW comp.) to a Cryogenic Sapphire Oscillator (CSO), itself locked to a H-maser on long time scales. Finally the frequency of the CSO is measured by the three caesium atomic fountains FO1, FO2 and FOM independently (see Methods).

We have built two OLCs[12,22,30] based on the transition at 698 nm between the ground state $^1S_0$ and the metastable state $^3P_0$ of neutral $^{87}$Sr atoms, with a natural linewidth of 1 mHz. The atoms are tightly confined in a few thousand wells of an optical lattice formed by a standing wave in an optical resonator. This strong confinement regime (Lamb-Dicke regime) dramatically reduces the motional effects. The lattice induces a light shift of several $10^5$ Hz on the energy levels, but, as anticipated in the initial proposal[9], we can control the differential shift at the millihertz level[22] when the lattice is operated at the magic wavelength. Together with the $J = 0$ nature of the clock states, this leads to a spectroscopy almost immune to fluctuations in trapping power or polarization. With consistent data accumulated over two years, we have evaluated the magic frequency cancelling the differential scalar polarizability: $f_{magic}$ = 368554725 (5) MHz. Since the optical lattice is the keystone of OLCs, we opted for two different trapping geometries[30] to test the robustness of the clocks against technical choices (Methods). The resonators forming the lattices allow trap depths as large as 5000 $E_r$, $E_r$ being the recoil energy of a strontium atom absorbing a lattice photon. This provides an important leverage to explore the residual effects of the lattice[22]. We could thus resolve the second order light shift, equal to 0.45(10) µHz/$E_r^2$ for a lattice polarization parallel to the quantization axis. These large depths, equivalent to temperatures of 0.8 mK, enable the capture of $10^4$ atoms in 500 ms directly from the first stage 2 mK magneto-optical trap based on the $^1S_0$–$^1P_1$ transition at 461 nm[12]. An extra cooling step is performed on the narrow line $^1S_0$–$^3P_1$, before the potential is ramped down adiabatically to 100 $E_r$. This leads to atomic temperatures of 1 µK in the strong confinement direction, and 10 µK in the transverse direction.

The large number of atoms probed simultaneously, together with the high Q-factor of the optical resonance, yields remarkable ultimate stabilities, $10^{-17}$ or below at one second if the Quantum Projection Noise (QPN) limit is reached. Steady work on ultrastable lasers[18-20,31,32] and interrogation sequences[17,33] has brought OLCs close to this limit[17-20] Here, the two atomic clouds are probed independently by pulses delivered by an ultrastable laser (spectral width <1 Hz). These pulses can be as long as 250 ms, resulting in Fourier-limited resonances of 3.2 Hz. Then, for each clock, the laser is locked to the narrow resonance via a digital integrator acting on a frequency shifter (Fig. 1). The difference between the corrections applied to the shifters gives the instantaneous frequency difference between the two clocks. Its relative stability is $3.0\times10^{-15}/\sqrt{\tau}$, reaching a resolution of $3\times10^{-17}$ after an integration time $\tau$ of 3 hours (Fig. 2).

The optical clocks feature an accuracy budget of respectively $u_1$ = $1.03\times10^{-16}$ and $u_2$ = $1.09\times10^{-16}$ detailed in the Table 1 and in the Methods. The main contribution to this budget, $7\times10^{-17}$, is the uncertainty on the Black-Body Radiation (BBR) shift[34] due to temperature inhomogeneities (~1K) in the environment surrounding the atoms. However, temperature-controlled environments can successfully handle this effect. Recently, the uncertainty on the atomic sensitivity to the BBR was dramatically reduced by experimental measurements[35-37], corroborated by theoretical calculations[38-40] (previous accuracy budgets of references 14, 15 and 16 are also reduced accordingly). To date, no effect seems in position to prevent OLCs accuracy from progressing by at least one order of magnitude.

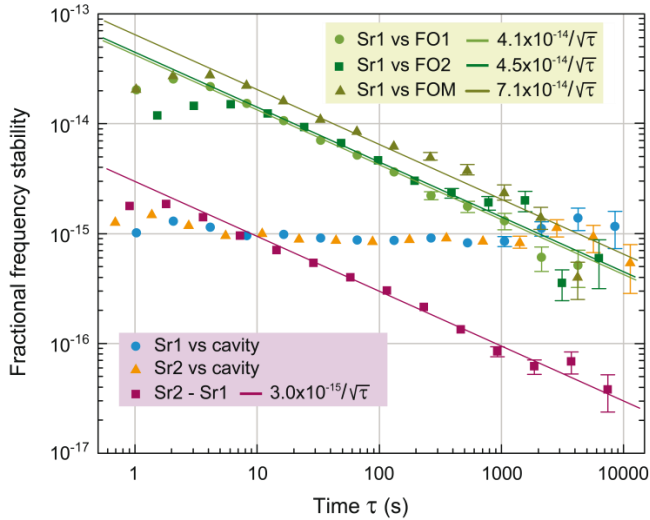

**Fig. 2: Relative frequency stabilities, for a single acquisition.** This Allan standard deviation gives the resolution achieved when comparing two oscillators for an integration time $\tau$. The blue circles and the orange triangles show the stability of the corrections applied to the frequency shifters going to each Sr clock (a predictable residual drift of the USC698 is removed). In the first seconds, these curves are dominated by the Dick effect (sampling of the high frequency noise of the clock laser by the atoms), here for a clock pulse of 110 ms and respective cycle times of 1 s (Sr1) and 0.7 s (Sr2). After 8 seconds, the fundamental thermal flicker noise of the USC698 becomes predominant, at the $8\times10^{-16}$ level. However, the direct comparison between the two devices (purple squares) shows a white noise behaviour scaling as $1/\sqrt{\tau}$. Its resolution reaches $3\times10^{-17}$ in 3 hours, equivalent to $2\times10^{-17}$ per clock. These optical-optical comparisons already beat by more than one order of magnitude the optical-microwave comparisons (green curves), dominated by the quantum projection noise of the fountains at $4.1\times10^{-14}$ at 1 s in the best case.

Comparing two clocks using the same atomic species is a crucial test to corroborate the fact that no systematic has been overlooked in their individual accuracy budget. This is the only case where the frequency ratio is known in advance: it is expected to be exactly 1 within the accuracy budget.

While this procedure is extensively validated on local[41] and global[2] scales for microwave clocks, agreement between identical optical clocks, with a total uncertainty surpassing the accuracy of the best Cs standards, was reached only for Al$^+$ ion clocks (ref 8, total uncertainty $2.7\times10^{-17}$) but had so far never been achieved for OLCs.

| Effect | Clock 1 (Sr1) | | Clock 2 (Sr2) | |
|---|---|---|---|---|
| | Corr. (mHz) | Uncertainty (mHz [in $10^{-17}$]) | Corr. (mHz) | Uncertainty (mHz [in $10^{-17}$]) |
| Quadratic Zeeman effect | 846 | 9 [2.1] | 558 | 6 [1.4] |
| Residual lattice 1$^{st}$ order LS | -21 | 5 [1.2] | 119 | 8 [1.9] |
| Lattice 2$^{nd}$ order LS | 0 | 3 [0.7] | -2 | 1.5 [0.3] |
| Optical amplifier spectrum | - | - | -110 | 15 [3.5] |
| Blackbody radiation<br>- uncertainty of temperature<br>- uncertainty of sensitivity | 2310 | 32 [7.5]<br>2 [0.5] | 2216 | 31 [7.2]<br>2 [0.5] |
| Density shift | -10 | 20 [4.6] | -2 | 22 [5.2] |
| Line pulling | 0 | 20 [4.7] | 0 | 20 [4.7] |
| Probe LS | 0 | 0.6 [0.15] | 0 | 0.6 [0.15] |
| **Total Sr** | **3125** | **44.1 [10.3]** | **2779** | **46.7 [10.9]** |

**Table 1: Accuracy budget** (LS=Light-Shift, Corr.= Correction)

We have conducted repeated high resolution comparisons between our two Sr clocks. Most acquisitions lasted several hours, ensuring a statistical uncertainty in the $10^{-17}$ range for each measurement, much lower than the systematic uncertainties. The frequencies of the two devices are in agreement, with a residual difference smaller than the systematic uncertainty (Fig. 3):

$$\Delta f_{Sr}/f_{Sr} = 1.2\times10^{-16} \pm 0.2\times10^{-16} \text{ (stat)} \pm 1.5\times10^{-16} \text{ (syst)}$$

This agreement was reached only after several unexpected shifts, due to patch charges inducing a DC Stark effect[30] and to the residual spectral background of lattice lights (Methods), had been identified and taken into account. This illustrates that direct comparisons between identical clocks, but with differences in the key features, are essential tests to support the exhaustiveness of the accuracy budget. Such shifts would have been undetectable with differential frequency measurements performed on a lone clock.

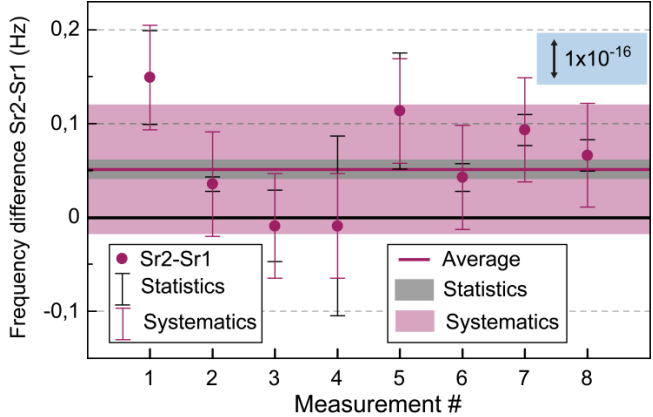

**Fig. 3: Direct comparison of the two strontium clocks, carried out over one week.** The acquisitions were carried out over one week, the resolution of the measurements (black error bars) is such that most of them are limited by systematics (purple error bars). The lattice frequencies were changed for the three last points to check the robustness of the results. The average frequency difference (solid purple line) between the two clocks, $1.2 \times 10^{-16}$, is compatible with zero within the systematic error bar (purple-shaded area, corresponding to the combined uncertainty of two uncorrelated clocks: $\sqrt{u_1^2 + u_2^2} = 1.5 \times 10^{-16}$). The grey-shaded area corresponds to the stability of the cumulated data. The $\chi^2$ of this average (0.55) is smaller than 1, which is typical of systematics fluctuating less than statistics.

Continuity at the best possible level through revisions of the definition of units is key. It is notably the case for timescales, used through decades for advanced scientific applications such as monitoring of pulsar timing or Earth rotation. In the case of OLCs, ascertaining the continuity requires absolute frequency measurements at the limit of the microwave-defined second. At LNE-SYRTE, the strontium clock frequency was measured extensively with a unique ensemble of three state-of-the-art caesium fountains[41]. This ensemble is operated quasi-continuously, with strong connections to the international realization of the SI second[2].

The titanium-sapphire laser based frequency comb[3,4] at the heart of our measurement chain (Fig. 1) connects two frequency domains separated by 5 orders of magnitude. The comb is phase-locked with a bandwidth of 400 kHz to a laser referenced to an ultrastable auxiliary cavity at 1062 nm (USC1062) with a thermal noise floor at $4 \times 10^{-16}$. The comb reaches therefore the narrow linewidth regime, with teeth spectrally narrower than 1 Hz. The comb measures the frequency ratio between two ultrastable oscillators: in the optical domain, the Sr clock laser at 698 nm (referenced to the ultrastable cavity USC698 exhibiting a thermal noise floor of $8 \times 10^{-16}$), and, in the microwave domain, a Cryogenic Sapphire Oscillator (CSO, at 11.980 GHz, exhibiting a short term stability in the $10^{-15}$ range, and locked to a H-maser to avoid long term drifts). By dividing the Sr laser frequency, the comb generates an ultrastable microwave signal referenced to the optical domain[42,43], which is then compared to the CSO. This approach covers an important aspect of a possible redefinition of the second: this microwave signal could feed directly any conventional or satellite system in need of a frequency reference, the existing infrastructure would thus need no modification. At the same time, the absolute frequency of the CSO is simultaneously and independently measured by the three Cs fountains, while, at the other end of the chain, the frequency difference between the Sr clock transition and the USC698 is given by the corrections applied to the frequency shifter. These measurements are combined to provide the absolute frequency of Sr against each fountain. We demonstrate the best frequency stability between an optical clock and a microwave clock ever reported to the best of our knowledge, at $4 \times 10^{-14}$ at one second (Fig. 2). This number is solely limited by the QPN of the fountains, which constitutes a fundamental limitation. On 11 non-consecutive days, the absolute frequency of $^{87}$Sr was measured against the three caesium fountains simultaneously (Fig. 4a). The independent averages are shown in Table 2. The large number of acquisitions ensured a statistical uncertainty lower than the systematic effects for each fountain. Furthermore, these three measurements agree within the systematic error bar, showing the consistency between fountains. The total average (Fig. 4b) yields:

$f_{Sr}$ = 429228004229873.10 ± 0.05 (stat) ± 0.12 (syst) Hz

Since a fundamental progress of the microwave clocks does no longer seem possible, we anticipate that this result borders on the ultimate accuracy at which any optical frequency can be measured in units of caesium-defined Hz.

Five groups worldwide have reported absolute frequency measurements of the $^{87}$Sr clock transition in the last 7 years (Fig. 4c). In this context, the data points we present strengthen the overall consistency, with an increased resolution at $3.1 \times 10^{-16}$. Our measurements, spanning over 6 months, enable us to track a change of the Sr/Cs frequency ratio due to possible variations of fundamental constants, which would reveal physics beyond the standard model. All the measured strontium frequencies are fitted by a linear drift ($-3.3 \times 10^{-16} \pm 3.0 \times 10^{-16}$/year) superimposed on a yearly sinusoidal contribution ($1.2 \times 10^{-16} \pm 4.4 \times 10^{-16}$) modelling a hypothetical coupling of the constants with the sun's gravitational potential that would reflect a violation of the Local Position Invariance principle (Fig. 4c). This fit is still compatible with no variation of the constants, but with an 8-fold increased resolution compared to the previous measurements[44]. These laboratory tests of fundamental physics, involving optical-microwave as well as optical-optical[5] and microwave-microwave[45] frequency ratios are an important alternative to astronomical evidences to constraint theories in the ongoing quest for a unified description of gravitation and quantum mechanics. They are all the more topical in view of recent results reporting a spatial variation of the fine structure constant from observations of quasars[46].

| Fountain | Measured strontium frequency (Hz) | Total statistics (mHz [in $10^{-16}$]) | Syst. (mHz [in $10^{-16}$]) Cs | Syst. (mHz [in $10^{-16}$]) Sr |
|---|---|---|---|---|
| FO1 | 429228004229873.11 | 73 [1.7] | 185 [4.3] | |
| FO2 | 429228004229873.07 | 99 [2.3] | 112 [2.6] | 47 [1.1] |
| FOM | 429228004229873.22 | 99 [2.3] | 344 [8.0] | |
| Average | 429228004229873.10 | 52 [1.2] | ~112 [2.6] | 47 [1.1] |
| | | | 121 [2.8] | |
| | | | 132 [3.1] | |

**Table 2: Absolute strontium frequency measurements** (Syst.= Systematics). These three measurements are in excellent agreement according to the respective error bar of each fountain. The final systematic error bar is mostly given by the best of the three fountains (FO2).

Our architecture of five state-of-the-art clocks, two strontium OLCs and three caesium fountains, are ticking with an unprecedented consistency. The link between the optical frequency domain and the microwave second is demonstrated to be reproducible and henceforth hindered by the hard limits the fountains are facing in accuracy ($2.6 \times 10^{-16}$) and in stability ($4 \times 10^{-14}$ at 1s due to the QPN). Direct comparisons between the two OLCs were necessary to show a level of agreement ($1.2 \times 10^{-16} \pm 1.5 \times 10^{-16}$) better than what can be resolved by three of the best caesium fountains. Our study shows that strontium optical lattice clocks provide a reproducible frequency with a trustworthy connection to the current SI second, and therefore to the various time scales established in the last 50 years. Several steps are nevertheless still required in order to demonstrate that OLCs are suited to a new definition of the second. Their accuracy continues to progress swiftly since the initial proposal[9], it is therefore necessary to further explore their fundamental limits, instead of choosing a definition that would be rapidly overrun. Ongoing efforts aiming at controlling the BBR shift by means of temperature-regulated environments raise the prospect of accuracies at $1 \times 10^{-17}$ or below for OLCs in a near future, thus possibly matching the performances of single ion clocks, which have taken the lead in terms of control of the systematics[5-8].

Finally, further comparisons, on a larger scale and between optical clocks designed and built in different institutes, are required to fully explore the potential of the various approaches. This process is crucial in order to confirm that all systematics have been comprehended, irrespective of experimental choices or local environment considerations. As an illustration, long term comparisons between state-of-the-art caesium fountains sometimes showed discrepancies exceeding the individual accuracy budgets in the past. This highlights the complexity of establishing a universal time unit. Any redefinition of the SI second would require an international consensus through due process, under the supervision of the CIPM. The ongoing development of coherent long-distance fibre links and the associated dissemination of ultra-pure optical carrier let us contemplate the possibility of an all-optical clocks network. These links will soon enable comparisons between remote optical clocks, with a level of performance taking full advantage of the record stabilities recently permitted by OLCs[20], thus allowing the exploration of clock's physics at the <$10^{-18}$ level. Such an extended network of clocks opens up new horizons for fundamental physics, such as testing of Einstein's equivalence principle and of the gravitational redshift. Further, clock-based geodesy with ultimate precision already raises new prospects in Earth Science[47,48].

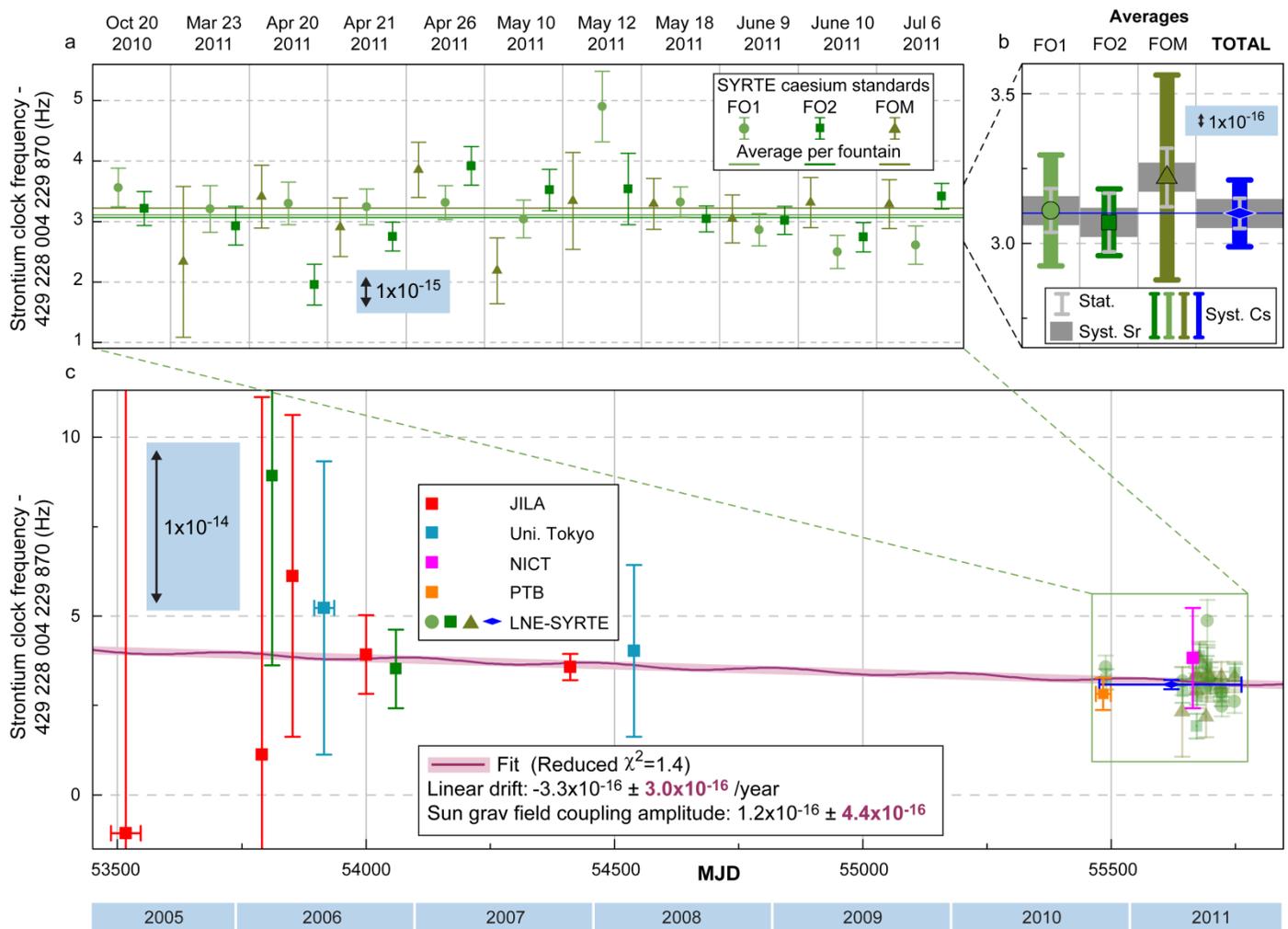

**Fig. 4: Absolute frequency measurements of the [87]Sr clock frequency. a.** The 11 LNE-SYRTE acquisitions, equivalent to 30 hours of measurement, are scattered over 6 months. For each acquisition, the strontium clocks are compared to the 3 LNE-SYRTE fountains (FO1, FO2, FOM). The total error bars represented with each point are dominated by the statistical uncertainty. **b**. Average per fountain of the points shown in **a**. For each fountain, the statistical error bar (thin lines) is smaller than the systematic uncertainty (thick lines), thus giving 3 independent measurements of the [87]Sr clock frequency limited by the fountain accuracy. These 3 values are in excellent agreement, reinforcing the confidence in our final average (blue diamond). **c**, Worldwide history of the [87]Sr clock frequency measurements reported by 5 different groups with their total error bar (ref 44 and references therein, ref 16, 49 and 50 modified accordingly to the BBR sensitivity from ref 36,40). Data points corresponding to **a** are shown in light green, the blue diamond represent the final average of **b**. The measurement campaign presented here enables us to improve by a factor 8 the bound on the time and spatial variations of the Sr/Cs frequency ratio.

## Methods

**Geometry of the lattices:** In order to demonstrate the level of control on the trapping effects, the two cavities forming the optical lattices have been designed in two different ways. The waist of the first lattice is 90 μm, and the intracavity power is 7 W, ensuring a maximum depth of 1000 $E_r$. With a waist of 56 μm and an intracavity power of 14 W, the second clock features potentials as deep as 5000 $E_r$. The magic frequency $f_{magic}$ = 368554725 (5) MHz ensures only the cancellation of the scalar part of the differential polarizability; a tensor correction[22] depending on the angle $\alpha$ between the lattice polarization and the quantization axis defined along the magnetic field can shift noticeably the effective magic point (from $f_{magic}$−268 MHz to $f_{magic}$+134 MHz). This angle also defines the sensitivity of the clocks to residual lattice polarization fluctuations. In the second lattice cavity, the absence of polarization selective element makes it difficult to ensure a perfectly linear polarization: $\alpha$ is thus chosen to minimize the sensitivity of this clock ($\alpha$ = 12 ± 5°). In the first clock cavity, on the other hand, an intracavity mirror selects a highly stable linear polarization (> 99.9% in field), $\alpha$ is adjusted to almost maximize the sensitivity ($\alpha$ = 51.5 ± 1°).

**Zeeman shift:** In order to control the influence of the magnetic field on the clock frequency, we optically pump the atoms alternately into the extreme Zeeman ground states $m_F$ = 9/2 or −9/2, the axis being defined along a bias field of 1 Gauss used to display the Zeeman structure. We then interleave measurements on the transitions $|^1S_0, F = 9/2, m_F = 9/2\rangle \rightarrow |^3P_0, F' = 9/2, m_F' = 9/2\rangle$ and $|^1S_0, F = 9/2, m_F = -9/2\rangle \rightarrow |^3P_0, F' = 9/2, m_F' = -9/2\rangle$, with no change of the quantum numbers describing the motion of the atoms in the lattice. The average of the two measured frequencies leads to a cancelation of the linear Zeeman shift and of the vector light shift. The difference of the two frequencies gives a real time monitoring of the magnetic field $B$ and of the ellipticity of the lattice polarization[22]. On long time-scales ($10^4$ s), $B$ is constant at a level of 0.1%, thus the uncertainty on the second order Zeeman shift correction ($2\times10^{-17}$) is limited by the knowledge of the sensitivity coefficient at the 1% level.

The experiment was designed to avoid any bias on the clock frequency due to a perturbing magnetic field that would be sequentially switched between two different values when the measurements are performed either on the $9/2 \rightarrow 9/2$ or on the $-9/2 \rightarrow -9/2$ transitions. No component producing a magnetic field is switched differently between the $m_F$ = 9/2 and the $m_F$ = −9/2 configurations. The ellipticity of the optical pumping light polarization is rather controlled by addressing alternately different AOMs.

Transients due to the switching of the MOT coils and of the orientation field were checked to take place in less than a few 100 μs, affecting the atomic populations at an insignificant level of $10^{-6}$. More generally, the clocks were shown to be immune to other transients by introducing various delays prior to the clock pulse. No frequency shift could be resolved and no loss of stability was observed.

**Motional and lattice induced effects:** The motional effects are dramatically suppressed in an OLC by the strong confinement of the atoms. The main residual motional effect is the pulling of the clock resonance by the transitions corresponding to a change of motional quantum number. Together with the pulling by the neighbouring Zeeman transitions (e.g. $m_F$ = 7/2 → $m_F'$ = 7/2 and $m_F$ = −7/2 → $m_F'$ = −7/2), the uncertainty due to this effect is at most $5\times10^{-17}$. A possible sliding of the lattice cavity, and therefore of the atoms, during the spectroscopy phase must also be taken into account. This effect was checked to happen at a maximum velocity of $v = 6\times10^{-9}$ m/s, inducing frequency shifts smaller than $2\times10^{-17}$ in relative value.

The lattices are operated close to their respective magic configuration, and their frequencies are stabilised via a transfer cavity to the $|^1S_0\rangle \rightarrow |^3P_1\rangle$ transition. The residual first order light shifts induced on the atoms are controlled at a level better than $2\times10^{-17}$, whereas the second order light shifts will not compromise the accuracy of the OLCs down to below $1\times10^{-17}$.

**Clock laser:** The power of the clock pulse is typically a few nW, and the light shift induced by the far-detuned coupling of the clock states with the other atomic levels was shown to be limited to $1.9 \times 10^{-17}$. The pulse is delivered by an AOM, the initial and final phase transients are limited to $\pi$ in less than 20 µs, inducing a negligible fractional excitation of $10^{-7}$ on the atoms. Phase drifts due to slow thermal effects in the AOM in the course of the pulse are observed, but they scale down with the RF power that is used. In our experimental configuration, we use 8 dBm of RF power, and we characterized that this induces a frequency bias of at most $5 \times 10^{-18}$.

**Collisional effects:** Atomic density effects in fermionic lattice clocks have drawn the attention in the last years, and it was demonstrated they could be as high as several $10^{-16}$ for $10^{11}$ atoms/cm$^3$ in some configurations (ref 10 and references therein, 25). The atomic density in our lattices is considerably lower ($10^9$ atoms/cm$^3$), and our measurements have confirmed the effect was not a concern at this stage: by measuring differentially the clock frequency in a nominal density configuration (~5 atoms/site) and in a low density configuration (~0.5 atoms/site), we have put an upper bound of $5 \times 10^{-17}$ on this effect (Fig. 5). According to the density ratio compared to ref 10, the effect is even projected to be less than $10^{-17}$.

**Black-body radiation shift:** The frequency response of strontium atoms to the BBR was measured recently 36 and validated theoretically 12, reducing considerably the corresponding uncertainty to $0.5 \times 10^{-17}$. The uncertainty on the BBR shift due to the imperfect control of the effective temperature experienced by the atoms is presently the main limitation to the clock accuracy budget. Temperature sensors were distributed around each of the vacuum chambers, and inhomogeneities of 1 K around the temperature of the first chamber (301 K) and of the second chamber (298 K) were observed in the course of the integrations. This results in a −2310 ± 32 mHz (resp. −2216 ± 31 mHz) shift on the Sr1 (resp. Sr2) clock.

The thermal radiation of the ovens can possibly affect the atoms, but it is considerably clipped by narrow differential pumping tubes interleaved between the ovens and the atoms. The residual effect was estimated to be at most 3 mHz. This is therefore negligible in relation to the above-mentioned uncertainties around 30 mHz ($7.5 \times 10^{-17}$).

**DC Start shift:** Static charges were trapped on the surface of the mirrors under vacuum forming the short 60 mm lattice cavity for the Sr2 clock. This induced a DC Stark shift on the atoms inducing a slowly decaying frequency shift on the order of $10^{-13}$. The charges were eventually removed from the coatings by photoelectric effect when shining UV light through an optical access of this chamber, the residual effect was characterized to be compatible with 0, with an uncertainty of $1.5 \times 10^{-18}$ (ref 30).

**Optical amplifier spectrum:** The residual spontaneous emission background of the Tapered Amplifiers (TAs) generating the light for the lattices was shown to induce a fluctuating discrepancy on the order of $10^{-15}$ between the two clocks. We carried out tests with various TAs and with different ways to control the optical power in the lattices (feedback on the current of the TA, feedback on an acousto-optic modulator or on a setup liquid crystal waveplate + polarization beamsplitter interleaved between the TA and the lattice cavity), and they yielded very different results for the two clocks.

The Sr1 frequency was demonstrated to be immune to the TA choice or to the way to control the lattice depth, we attribute this feature to the geometry of the lattice cavity and particularly to the polarization purity.

On the other hand, the Sr2 provided discrepant results. The unwanted shift induced by a given TA for the clock operated at a given depth was calibrated by differential measurements between two configurations: one configuration with the lattice run with light stemming from the TA, the other configuration with the lattice run with light stemming from the ECDL seeding the TA in the first configuration. The TA is therefore completely bypassed during the spectroscopy phase in the second configuration. The shift was calibrated to be 110 ± 15 mHz at a lattice depth of 100 $E_r$, introducing therefore a 3.5×10$^{-17}$ additional uncertainty in the accuracy budget of Sr2.

**Contribution of the absolute frequency measurement chain:** The relative height of the different clocks was measured with a precision better than a few 10 cm, thus inducing an uncertainty of at most a few 10$^{-17}$ on the comparison. The accuracy of the compensated fiber link transferring the signal of the CSO has also been evaluated to be at the 10$^{-17}$ level. Finally, the measurement of the frequency of the same cavity (USC1062) simultaneously with the titanium-sapphire laser based comb and with two erbium-fiber based combs showed that the uncertainty of the frequency transfer by the combs is at most 3×10$^{-17}$.

**Remark:** Except for the BBR contribution, all the uncertainties have been conservatively evaluated. Once temperature-controlled environments are in place in state-of-the-art OLCs, there seems to be no effect prohibiting their accuracy from progressing by at least one order of magnitude.

**Acknowledgments:** SYRTE is UMR CNRS 8630. We acknowledge funding by LNE, CNES, IFRAF, DGA, EMRP project T1 J2.1 OCS, ESA project SOC1, FP7 project SOC2. We thank Christophe Salomon and Thomas Udem for fruitful discussions on the manuscript.

**Author Contributions:** All authors contributed extensively to the work presented in this paper.